\documentclass[preprint,12pt,3p]{elsarticle}

\usepackage{amssymb}

\journal{}

\begin{document}

\begin{frontmatter}

\title{Noncommutative quantum Hall effect in graphene}

\author[label1]{Willien O. Santos}
\address[label1]{Colegiado de Fisica, Universidade Federal do Reconcavo
da Bahia, 45300-000, Amargosa, BA, Brazil}
\ead{willien@ufrb.edu.br}

\cortext[cor1]{Corresponding author}

\author[label2]{Guilherme M. A. Almeida\corref{cor1}}
\address[label2]{Instituto de Fisica, Universidade Federal de Alagoas, 57072-970, Maceio, AL, Brazil}
\ead{gmaalmeidaphys@gmail.com}

\author[label4,label5]{Andre M. C. Souza}
\address[label4]{Departamento de Fisica, Universidade Federal de Sergipe, 49100-000, Sao Cristovao, SE, Brazil}
\address[label5]{Department of Physics, University of Central Florida, Orlando, FL 32816, USA}
\ead{amcsouza@ufs.br}

\begin{abstract}
We discuss the quantum Hall effect on a single-layer graphene in the framework of noncommutative (NC) phase space. We find 
it induces
a shift in the Hall resistivity. Furthermore, comparison with experimental data reveals an upper bound on the magnitude of the momentum NC parameter $\eta$ in about $\sqrt{\eta}\leq 2.5 \, \mathrm{eV}/c$.
\end{abstract}


\end{frontmatter}


\section{Introduction}

Since put forward several years ago \cite{snyder47}, the idea of space-time noncommutativity 
has drawn quite
a lot of interest in a wide range of areas \cite{douglasrev}
Formerly backed up by sound arguments from string theory \cite{connes98, seiberg99} -- in which noncommutative (NC) space effectively 
comes out 
as a background magnetic field in $D$-brane theory -- and high-energy physics \cite{douglasrev,carroll01},
NC physics
would also soon find support in quantum mechanics \cite{chaichian01, gamboa01}.

Naturally, the primary goal is to unveil observable traces 
of NC effects in more accessible experimental setups \cite{R26, zhang04, zhang12, liang14, santos17}
and set boundaries for their characteristic length and energy scales. 
In this context, various well-established subjects have been covered such as, to name a few,
the hydrogen
atom \cite{chaichian01}, 
quantum Hall physics \cite{R16,R17,R18,R19,R20,R21,R22}
anomalous Zeeman effect \cite{R15, R14}, 
Aharonov-Bohm\cite{R26,R24,R25}, and
Aharonov-Casher \cite{R27,R28,R29} effects. 
%
Quantum information theory was also inspected on the NC phase-space framework \cite{bernardini13, bastos13}

Quite recently, the so-called
graphene has been increasingly addressed, 
with most of works focusing on the implications of 
noncommutativity in its energy spectrum \cite{R37,R34,R36,R35}.
There is, however, no proper study
on how quantum Hall effects on graphene
would stand out in the NC phase-space configuration.
%
%
Given the non-stopping progress of such 
class of materiais on both theoretical \cite{castronetorev} and practical \cite{yurev} sides
and the relevance of studying the
phenomenological implications of noncommutativity in the condensed-matter domain,  
in this work we investigate how noncommuting momentum coordinates
affect the Hall conductivity/resistivity in graphene and, alongside experimental data found in \cite{R46}, use it to 
estimate an upper bound in the scale of the 
associated NC parameter, namely $\eta \leq (2.5 \, \mathrm{eV}/c)^{2}$,
which sets the energy scale where 
noncommutativity effects are likely to be measured.

\section{NC graphene in a magnetic field}

In NC phase space, the coordinates $\widehat{x}_{i}$ and momentum operators $\widehat{p}_{i}$ satisfy \cite{R30}
\begin{equation}
\begin{array}{l}
\left [\widehat{x}_{i},\widehat{x}_{j} \right ]=i\theta_{ij}, \\
\left [\widehat{x}_{i},\widehat{p}_{j}\right ]=i{\hbar}(\delta_{ij}+\frac{\theta_{ia}\eta_{aj}}{4\hbar^{2}}),\\
\left [\widehat{p}_{i},\widehat{p}_{j}\right ]=i\eta_{ij},
\end{array}
\label{eq.1}
\end{equation}
where $\theta_{ij}=\epsilon_{ijk}\theta_{k}/2$ and $\eta_{ij}=\epsilon_{ijk}\eta_{k}/2$ are antisymmetric constant matrices with dimensions of $(\mathrm{length})^{2}$ and $(\mathrm{momentum})^{2}$, respectively. For easiness,  herein we consider $\vec{\theta} = (0,0,\theta)$ and
$\vec{\eta} = (0,0,\eta)$.

The commutation relations (\ref{eq.1}) can be explicitly implemented by means of coordinate transformations\cite{R30},
\begin{equation}
\widehat{x}_{i}=x_i-\frac{1}{2\hbar}\theta_{ij}p_j,\ \ 
\widehat{p}_{i}=p_i+\frac{1}{2\hbar}\eta_{ij}x_{j},
\label{eq.2}
\end{equation}
where variables $x_{i}$ and $p_{i}$ satisfy the usual canonical commutation relations 
\begin{equation}
\begin{array}{l}
\left [x_{i}, x_{j} \right ]=0, \\
\left [x_{i}, p_{j}\right ]=i{\hbar}\delta_{ij},\\
\left [p_{i}, p_{j}\right ]=0.
\end{array}
\label{eq.3}
\end{equation}

Our starting point is the Dirac equation for a free massless particle
\begin{equation}
i\hbar\frac{\partial\psi}{\partial t}=\widehat{H}_{D}\psi,
\label{eq.4}
\end{equation}
where $\widehat{H}_{D}$ is the NC massless Dirac-like Hamiltonian that describes the electron around Dirac points $K$ and $K'$ at the corners of the Brillouin zone\cite{R38}, thus reading
\begin{eqnarray}
\widehat{H}_{D}=\left( \begin{array}{cl}
\widehat{H}^{K} & 0 \\
0 & \widehat{H}^{K'}
\end{array}\right) =
v_{f}\left(\begin{array}{cl}
\vec{\sigma}\cdot\widehat{\vec{p}} & 0 \\
0 & \vec{\sigma}^*\cdot\widehat{\vec{p}}
\end{array}\right),
\label{eq.5}
\end{eqnarray}
where $v_{f}$ is the Fermi velocity playing the role of speed of light in vacuum, $\widehat{\vec{p}}=(\widehat{p}_{x},\widehat{p}_{y})$ is the two-dimensional momentum operator,
and $\vec{\sigma}=(\sigma_{x}, \sigma_{y}, \sigma_{z})$ are the usual Pauli matrices acting on states of both sub-lattices $A$ and $B$. 
The four-component wave function corresponding to Hamiltonian (\ref{eq.5}) is 
\begin{equation}
\psi = \left( \begin{array}{l}
\psi^{K} \\
\psi^{K'}
\end{array}\right),
\label{eq.6}
\end{equation}  
with 
\begin{equation}
\psi^{K} = \left( \begin{array}{l}
\psi^{K}_{A} \\
\psi^{K}_{B}
\end{array}\right), \ \psi^{K'} =\left( \begin{array}{l}
\psi^{K'}_{A} \\
\psi^{K'}_{B}
\end{array}\right),
\label{eq.7}
\end{equation}
where $\psi^{K}_{A}$ and $\psi^{K}_{B}$ ($\psi^{K'}_{A}$ and $\psi^{K'}_{B}$) are wave-function envelops at $A$ and $B$ for valley $K$ ($K'$).

In order to describe a free electron subjected to a 
uniform magnetic field $\vec{B}$ in the positive $z$-direction, one needs to replace the gauge invariant kinetic momentum through \cite{R39}
\begin{equation}
\widehat{\vec{p}}\rightarrow\widehat{\vec{\Pi}}=\widehat{\vec{p}}+\frac{e}{c}\widehat{\vec{A}} = (1-\frac{eB\theta}{8c\hbar})\vec{p}+\frac{e}{c}(1-\frac{\eta c}{2eB\hbar} )\vec{A},
\label{eq.8}
\end{equation}
where $\widehat{\vec{A}}$ is the NC vector potential given by the Coulomb gauge, $\widehat{\vec{A}}=(-B\widehat{y}/2, B\widehat{x}/2)$. 
As pointed out in Ref. \cite{R40} the factor containing $\theta$ in eq.~(\ref{eq.8}) breaks the gauge symmetry, what makes 
the electron velocity not gauge invariant in NC phase space. However, by taking $\theta = 0$, 
gauge symmetry is preserved. 
Indeed, by only letting spatial coordinates to commute, that is   
$\theta =0$ and $\eta\neq 0$, the momentum variables $\widehat{\Pi}_{i}$ satisfy 
the non-canonical commutation relation
\begin{equation}
\left [\widehat{\Pi}_{x},\widehat{\Pi}_{y} \right ]=\frac{-i\tilde{\hbar}^{2}}{\tilde{l}^{2}_{B}},
\label{eq.9}
\end{equation}
where
\begin{equation}
 \tilde{l}_{B}=\sqrt{\frac{\lambda\hbar c}{eB}}=\sqrt{\lambda}l_{B}, \tilde{\hbar}=\lambda\hbar, \lambda=1-\frac{\eta c}{2eB\hbar}.
\label{eq.10}
\end{equation}

We now work out the eigenvalues of Eq~(\ref{eq.5}). In order to do so, 
we introduce NC ladder operators of the form
\begin{equation}
\widehat{a}=\frac{\tilde{l}_{B}}{\sqrt{2}\tilde{\hbar}}\left(\widehat{\Pi}_{x}-i\widehat{\Pi}_{y} \right), \ \widehat{a}^{\dagger}=\frac{\tilde{l}_{B}}{\sqrt{2}\tilde{\hbar}}\left(\widehat{\Pi}_{x}+i\widehat{\Pi}_{y} \right),
\label{eq.11}
\end{equation}
which obey $[\widehat{a}, \widehat{a}^{\dagger}]=1$. Likewise, we shall write
\begin{equation}
\widehat{\Pi}_{x}=\frac{\tilde{\hbar}}{\sqrt{2}\tilde{l}_{B}}\left(\widehat{a}+ \widehat{a}^{\dagger}\right), \ \widehat{\Pi}_{y}=\frac{\tilde{\hbar}}{i\sqrt{2}\tilde{l}_{B}}\left(-\widehat{a}+ \widehat{a}^{\dagger}\right).
\label{eq.12}
\end{equation}

In terms of the ladder operators introduced above, Hamiltonian (\ref{eq.5}) becomes
\begin{equation}
\widehat{H}^{K}=\tilde{\omega}\hbar\left( \begin{array}{ll}
0 & \widehat{a} \\
\widehat{a}^{\dagger} & 0
\end{array}\right),
\label{eq.13}
\end{equation}
where $\tilde{\omega}=\sqrt{2}v_{f}/\tilde{l}_{B}$ is the relativistic cyclotron frequency.
The eigenvalues of Hamiltonian (\ref{eq.13}) are obtained solving
\begin{equation}
\widehat{H}^{K}\psi^{K}=\widehat{E}^{K}\psi^{K},
\label{eq.14}
\end{equation}
which leads to
\begin{eqnarray}
\tilde{\omega}\hbar\widehat{a}\psi^{K}_{B}&=\widehat{E}^{K}\psi^{K}_{A}, \\ \tilde{\omega}\hbar\widehat{a}^{\dagger}\psi^{K}_{A}&=\widehat{E}^{K}\psi^{K}_{B},
\label{eq.15}
\end{eqnarray}
and then
\begin{equation}
\widehat{a}^{\dagger}\widehat{a}\psi^{K}_{B}=\left(\frac{\widehat{E}^{K}}{\hbar\tilde{\omega}} \right) \psi^{K}_{B}.
\label{eq.16}
\end{equation}
The state corresponding to spinor component $\psi^{K}_{B}$ becomes
\begin{equation}
\psi^{K}_{B}\equiv \vert n\rangle = \frac{(\widehat{a}^{\dagger})^{n}}{\sqrt{n!}}\vert  0\rangle ,
\label{eq.17}
\end{equation}
with $n=0, 1, 2,...$ and, as usual, $\widehat{a}^{\dagger}\vert  n\rangle =\sqrt{n+1}\vert n+1\rangle$, 
$\widehat{a}\vert  n\rangle =\sqrt{n}\vert n-1\rangle$. 

Thus, Eq.~(\ref{eq.16}) can be solved to obtain the 
corresponding NC Landau level (LL) energy spectrum.
\begin{equation}
\widehat{E}^{K}=\pm\frac{\hbar v_{f}}{\tilde{l}_{B}}\sqrt{2n},
\label{eq.18}
\end{equation}
where positive values correspond to electrons (conduction band), 
and negative values stand for holes (valence band).

\section{NC quantum Hall effect in graphene}
Classically, LL degeneracy is related to the center of the cyclotron motion (guiding center), which is a constant of motion\cite{R39}. 
Consequently, the NC position operator can be 
split,
\begin{equation}
\widehat{\vec{r}}=\widehat{\vec{R}}+\widehat{\vec{\eta}},
\label{eq.19}
\end{equation}
into the NC guiding center $\widehat{\vec{R}}=(\widehat{X}, \widehat{Y})$ and
the corresponding NC cyclotron variables $\widehat{\vec{\eta}}=(\widehat{\eta}_{x}, \widehat{\eta}_{y})$. A relationship between $\widehat{\vec{\eta}}$ and 
$\widehat{\vec{\Pi}}$ can be obtained through
\begin{equation}
\frac{d\widehat{\vec{\Pi}}}{d t}=-\frac{e}{c}\frac{d\widehat{\vec{\eta}}}{dt}\times\vec{B},
\label{eq.20}
\end{equation}
which yields 
\begin{equation}
\widehat{\eta}_{x}=\frac{c\widehat{\Pi}_{y}}{eB}, \ \widehat{\eta}_{y}=-\frac{c\widehat{\Pi}_{x}}{eB},
\label{eq.21}
\end{equation}
and, using Eq.~(\ref{eq.9}), it is straightforward to check that 
\begin{equation}
\left[\widehat{\eta}_{x}, \widehat{\eta}_{y} \right] =-i\tilde{l}_{B}^{2}.
\label{eq.22}
\end{equation}
Thereby, Eqs.~(\ref{eq.19}) and (\ref{eq.22}) brings about the commutation relation
\begin{eqnarray}
\left[\widehat{X}, \widehat{Y} \right] &=& \left[ \widehat{x}, \widehat{y}\right] -\left[ \widehat{x}, \widehat{\eta}_{y}\right]-\left[ \widehat{\eta}_{x}, \widehat{y}\right]+\left[\widehat{\eta}_{x}, \widehat{\eta}_{y} \right]\cr &\simeq & -\left[\widehat{\eta}_{x}, \widehat{\eta}_{y} \right]=i\tilde{l}_{B}^{2},
\label{eq.23}
\end{eqnarray}
where $\eta c/2eB\hbar\ll 1$, $[\widehat{x}, \widehat{y}]=0$ and $[\widehat{x}, \widehat{\eta}_{y}]=[\widehat{\eta}_{x}, \widehat{y}]=-i\tilde{l}_{B}^{2}$. We are now able to use the above expression for 
counting down the number of quantum states. 
Indeed, that also indicates one may not measure both components of
$\widehat{\vec{R}}$ simultaneously, which leads to the 
quantized NC cyclotron radius
\begin{equation}
\widehat{r}_{n}=\frac{\vert\widehat{E}^{K}\vert}{eBv_{f}/c}=\sqrt{2n}\tilde{l}_{B}.
\label{eq.24}
\end{equation}
The area occupied by a quantized cyclotron orbit is thus 
\begin{equation}
\Delta\widehat{X}\Delta\widehat{Y}=\pi\widehat{r}_{n+1}^{2}-\pi\widehat{r}_{n}^{2}=2\pi\tilde{l}_{B}^{2},
\label{eq.25}
\end{equation}
and so the number of possible quantum states on a given surface $A$ is
\begin{equation}
\widehat{N}_{B}=\frac{A}{2\pi\tilde{l}_{B}^{2}}=\widehat{n}_{B}A,
\label{eq.26}
\end{equation}
with
\begin{equation}
\widehat{n}_{B}=\frac{eB}{2\pi\tilde{\hbar} c}
\label{eq.27}
\end{equation}
being the NC flux density, i.e., the magnetic field in units of the NC quantum flux $ 2\pi\tilde{\hbar}c/e $. 
Indeed, $\tilde{l}_{B}$ 
shall be understood noting that
surface $A$ contains a quanta of that flux. 
The ratio $g_{e}=\widehat{n}_{e}/\widehat{n}_{B}$
then characterizes the filling factor of different LLs. 

The standard quantum Hall effect occurs at low temperatures and large magnetic fields. 
In graphene, however, due to large energy scale between LLs 
the quantized Hall effect shall be displayed at room temperature \cite{R41,R42}. 
When the Fermi level crosses a given LL, the Hall conductivity increases for integer values of $g_{s}e^{2}/hc$, 
where $g_{s}$ is the degeneracy of the LL. In graphene, it takes $g_{s}=4$ and thus
one would expect the plateaus of Hall conductance 
to occur for integer values of $4e^{2}/hc$. 
However, the quantum Hall effect in graphene is anomalous and the quantization of the Hall conductance actually occurs for half-integer values of
\begin{equation}
\sigma_{H}=g_{e}\frac{e^{2}}{hc},
\label{eq.28}
\end{equation}
where the filling factor here is $g_{e}=4(\nu+1/2)=\pm 2, \pm 6, \pm 10,\ldots$, which is different from that of the conventional semiconductor quantum Hall effect, i.e, $g_{e}=\pm 4, \pm 8, \pm 12,\ldots$.

Now, in order to find the NC-induced shift in the Hall conductance and resistivity, we use 
the well-known Streda formula\cite{R43} at constant Fermi energy 
(see also\cite{R44,R45}) 
\begin{equation}
\sigma_{H}=e\frac{\partial n_{e}}{\partial B}\bigg|_{E=E_{f}}.
\label{eq.29}
\end{equation}
For a single-layer graphene on NC momentum space, the Hall conductivity reads
\begin{equation}
\widehat{\sigma}_{H}=g_{e}e\frac{\partial\tilde{n}_{e}}{\partial B}=g_{e}\frac{e^{2}}{hc}\left(1-\frac{\eta c}{eB\hbar} \right)\left(1-\frac{\eta c}{2eB\hbar} \right)^{-2}.
\label{eq.30}
\end{equation}
The electrical resistivity is thus $\widehat{\rho}_{H} = 1/\widehat{\sigma}_{H}$.
When $\eta=0$ then $\widehat{\rho}_{H}=\rho_{H}=hc/g_{e}e^{2}$ which is exactly the result of the usual commutative case.

\section{Phenomenological estimation for the momentum noncommutativity effect}

In the literature, there is a handful of 
studies aimed to provide upper bounds on the momentum NC parameter $\eta$. 
For instance, in\cite{R30} they discussed the gravitational quantum well in NC phase space and found 
$\eta\leq (1~\mathrm{meV}/c)^{2}$. 
It is crucial to note, though,
that such result depends heavily 
on an assumption about the configuration space NC parameter $ \theta $, 
which in this case was $\theta\leq (10~\mathrm{TeV})^{-2}$. 
%
%
In\cite{R40} the constraint $\eta\leq (2.26~\mu \mathrm{eV}/c)^{2}$ was attained
for the hyperfine transition in the hydrogen atom. 
Along those lines, a very stringent bound was obtained in\cite{R14} for the anomalous Zeeman effect in the hydrogen atom, 
namely $\eta\leq (0.34~\mu \mathrm{eV}/c)^{2}$, without setting a specific value for $\theta$.

Let us now carry out a similar analysis based on our findings [cf. Eq. ~(\ref{eq.30})].
Given that $\widehat{\rho}_{H} = 1/\widehat{\sigma}_{H}$, isolating the NC parameter $\eta$ gives us
\begin{equation}
\eta = \frac{2\hbar^{2}}{l_{B}^{2}}\left( -\vert\delta\widehat{\rho}_{H}\vert +\sqrt{\vert\delta\widehat{\rho}_{H}\vert(1+\vert\delta\widehat{\rho}_{H}\vert)}\right) ,
\label{eq.32}
\end{equation}
where $\delta\widehat{\rho}_{H}=(\widehat{\rho}_{H}-\rho_{H})/\rho_{H}$. A rather precise measurement of the Hall resistivity on a single-layer graphene can be found in \cite{R46}, $\delta\rho_{H}=15\cdot 10^{-6}$. Thereby, using the relative accuracy $\delta\widehat{\rho}_{H}\leq 15\cdot 10^{-6}$, for the external magnetic field $B\sim 10^{4} G$, we get
\begin{equation}
\eta\leq 1.8\cdot 10^{-44}~\mathrm{g}^{2}\mathrm{cm}^{2}\mathrm{s}^{-2}\Rightarrow\sqrt{\eta}\leq 2.5~ \mathrm{eV}/c.
\label{eq.33}
\end{equation}
We shall mention that the above value is of the same order of magnitude as the limit obtained in Ref.\cite{R36}.

\section{Conclusions}
We explored above the quantum Hall effect on a single-layer graphene in NC phase space, where momentum coordinates
did not commute, as accounted by the parameter $\eta$. 
We found out that this feature brings about a correction 
to the so-called Hall conductance/resistivity. 
Furthermore, comparison with experimental data \cite{R46} revealed that $\sqrt{\eta}\leq 2.5~ \mathrm{eV}/c$. 
Our result is higher 
than most of the upper limits found out elsewhere \cite{R14,R30,R40} in other physical frameworks. 
We would also like to point out that due to 
the underlying two-dimensional crystalline structure of graphene, the Hall effect resembles
the Landau problem, where noncommutativity effects are of the same order of magnitude (see, e.g., \cite{gamboa01-2}).

Our work extends previous results for graphene
in NC algebra settings\cite{R37,R34,R36,R35}
and, overall, also contributes to the endeavour of finding out 
phenomenological implications and physical signatures 
of noncommutativity in experimentally-feasible platforms
in the context of condensed-matter physics.

\section*{Acknowledgments}
This work was supported by CNPq. G.M.A.A. acknowledges funding over Grant No. 152722/2016-5.

\end{document}